\documentclass{aa}
\usepackage{graphicx}

\def\jh{\mbox{$\rm (J-H)$}}
\def\hk{\mbox{$\rm (H-K_S)$}}

\def\mMo{\mbox{$\rm (m-M)_O$}}
\def\ebv{\mbox{$\rm E(B-V)$}}
\def\ejh{\mbox{$\rm E(J-H)$}}
\def\rc{\mbox{$\rm R_{core}$}}
\def\rl{\mbox{$\rm R_{lim}$}}
\def\ms{\mbox{$\rm M_\odot$}}
\def\ds{\mbox{$\rm d_\odot$}}
\def\dgc{\mbox{$\rm d_{GC}$}}
\def\jj{\mbox{$\rm J$}}
\def\hh{\mbox{$\rm H$}}
\def\ks{\mbox{$\rm K_S$}}

\def\mobs{\mbox{$\rm m_{obs}$}}
\def\mtot{\mbox{$\rm m_{MS+PMS}$}}
\def\kms{\mbox{$\rm km\,s^{-1}$}}
\def\tr{\mbox{$\rm t_{relax}$}}
\def\tcr{\mbox{$\rm t_{cross}$}}

\begin{document}

\title{Mass functions and structure of the young open cluster NGC\,6611}

\author{C. Bonatto\inst{1}, J.F.C. Santos Jr.\inst{2} \and E. Bica\inst{1}}

\offprints{Ch. Bonatto - charles@if.ufrgs.br}

\institute{Universidade Federal do Rio Grande do Sul, Instituto de F\'\i sica, 
CP\,15051, Porto Alegre 91501-970, RS, Brazil\\
\mail{}
\and
Universidade Federal de Minas Gerais, ICEx, Departamento de F\'\i sica,
CP\,702, Belo Horizonte 30162-970, MG, Brazil
}

\date{Received --; accepted --}

\abstract{We use \jj, \hh\ and \ks\ 2MASS photometry to study colour-magnitude (CMDs) and
colour-colour diagrams, structure and mass distribution in the ionizing open cluster NGC\,6611. 
Reddening variation throughout the cluster region is taken into account followed by field-star 
decontamination of the CMDs. Decontamination is also applied to derive the density profile and 
luminosity functions in the core, halo and overall (whole cluster) regions. The field-star
decontamination showed that the lower limit of the main sequence (MS) occurs at $\rm\approx5\,\ms$. 
Based on 
the fraction of \ks\ excess stars in the colour-colour diagram we estimate an age of $1.3\pm0.3$\,Myr
which is consistent with the presence of a large number of pre-main sequence (PMS) stars. The 
distance from the Sun was estimated from known O\,V stars in the cluster area and the turn-on 
stars connecting the PMS and MS, resulting in $\rm\ds=1.8\pm0.5\,kpc$. The radial density 
distribution including MS and PMS stars is fitted by a King profile with a core radius 
$\rm\rc=0.70\pm0.08\,pc$. The cluster density profile merges into the background at a limiting radius 
$\rm\rl=6.5\pm0.5\,pc$. From the field-star subtracted luminosity functions we derive
the mass functions (MFs) in the form $\rm\phi(m)\propto m^{(-1+\chi)}$. In the halo and 
through the whole cluster the MFs have slopes $\rm\chi=1.52\pm0.13$ and $\rm\chi=1.45\pm0.12$,
respectively, thus slightly steeper than Salpeter's IMF. In the core the MF is flat, 
$\rm\chi=0.62\pm0.16$, indicating some degree of mass segregation since the cluster age is a factor
$\sim2$ larger than the relaxation time. Because of the very young age of NGC\,6611, part of 
this effect appears to be related to the molecular cloud-fragmentation process itself. We detect 
$362\pm120$ PMS stars. The total observed mass including detected MS (in the range
$\rm5-85\,\ms$) and PMS stars amounts to 
$\sim1\,600\,\ms$, thus more massive than the Trapezium cluster. Compared to older open clusters 
of different masses, the overall NGC\,6611 fits in the relations involving structural and dynamical 
parameters. However, the core is atypical in the sense that it looks like an old/dynamically evolved 
core. Again, part of this effect must be linked to formation processes.

\keywords{({\it Galaxy}:) open clusters and associations: individual: NGC\,6611; 
{\it Galaxy}: structure} }

\titlerunning{Structure of NGC\,6611}

\authorrunning{C. Bonatto, J.F.C. Santos Jr \and E. Bica}

\maketitle

\section{Introduction}
\label{intro}

The distribution of stellar masses at star-cluster birth is a complex process not yet fully 
understood. Recent observations of the stellar content of star-forming regions in molecular 
clouds, rich star clusters and the Galactic field suggest that the initial mass function (IMF) 
has similar properties in these very different environments (Kroupa \cite{Kroupa2002}). This 
seems to indicate that the initial distribution of stellar masses should depend only on the 
process of molecular cloud fragmentation, and that the fragmentation would have to produce 
similar IMFs despite very different initial conditions, a physical process which still lacks 
a better description (Kroupa \cite{Kroupa2002}). In this sense, detailed analysis of the spatial
structure and stellar-mass distribution in young star clusters may shed light on this issue.
A suitable target for this kind of study is the young open cluster NGC\,6611, which is responsible
for the ionization of the \ion{H}{ii} region Sh2-49 (Sharpless \cite{Sh59}), aka Gum\,83
(Gum \cite{Gum55}) or RCW\,165 (Rodgers, Campbell, \& Whiteoak \cite{RCW60}) as seen in the
optical, or W\,37 (Westerhout \cite{West58}) in the radio.

The open cluster NGC\,6611 (M\,16, Mel-198, Cr\,375, OCl\,54 - Alter, Ruprecht, \& Vanisek 
\cite{Alter1970}) consists of a system of early-type stars embedded in emission nebulosity 
(The Eagle Nebula) with spatially variable dust attenuation in the optical range. Structurally 
the region shows some nebulosity features, the so-called elephant trunks (the ``Pillars of 
Creation'', Hester, Scowen, Sankrit, et al. \cite{Hester96}), at the tips of which new-born 
stars start to be visible. 
  
The cluster is located at $\ell\approx17^\circ$ and $b\approx0.8^\circ$, in the Sagittarius 
spiral arm (Georgelin, \& Georgelin \cite{GG70}), a region presenting conspicuous dense dust 
clouds. The whole region is an active site where star-formation processes seem to occur 
giving birth to low to high-mass stars (Kroupa \cite{Kroupa2004}). Evidence 
has been presented in recent years that NGC\,6611 shows an extremely flat MF (e.g. Sagar \& 
Joshi \cite{Sagar79}; Sagar et al. \cite{Sagar86}; Hillenbrand et al. \cite{Hillenbrand93};
Massey et al. \cite{Massey95}). The distance of NGC\,6611 is dependent on the extinction law 
but several works constrain it to $\ds=2.3\pm0.3$\,kpc (Th\'e et al. \cite{twfw90}; Hillenbrand 
et al. \cite{Hillenbrand93}; Belikov et al \cite{Belikov99}). The cluster stars have been 
observed in evolutionary stages which are not reconcilable with a single age system (de Winter 
et al. \cite{w97}). Also, isochrone age determinations are ambiguous for such young clusters 
because of the lack of constraints in the CMD morphology. However, several studies show that 
the cluster is certainly younger than 10\,Myr with the youngest members still being formed (e.g. 
Hillenbrand et al. \cite{Hillenbrand93}; de Winter \cite{w97}; Belikov et al \cite{Belikov00}). 
NGC\,6611 presents a significant number of PMS stars. de Winter et al. (\cite{w97}) revised the 
subject of PMS membership and related issues by studying PMS candidates individually via photometry 
and spectroscopy. They find that the sample objects are not coeval, with an age spread of 
$\approx6$\,Myr being necessary to explain the cluster stellar content.

Our main goals with the present work are {\em (i)} carry out a detailed analysis of the spatial structure 
of NGC\,6611, {\em (ii)} analyze the spatial variation of the cluster MFs, {\em (iii)} derive properties 
of the MS and PMS stars, and {\em (iv)} infer on the dynamical state of this very young open cluster. For 
the sake of spatial and photometric uniformity, we employ \jj, \hh\ and \ks\ 
2MASS\footnote {The Two Micron All Sky Survey, All Sky data release (Skrutskie et al. \cite{2mass1997}), 
available at {\em http://www.ipac.caltech.edu/2mass/releases/allsky/}} photometry. The 2MASS Point 
Source Catalogue (PSC) is uniform reaching relatively faint magnitudes covering nearly all the sky, 
allowing a proper background definition even for clusters with large angular sizes (e.g.
Bonatto, Bica, \& Santos Jr. \cite{BBS2004}; Bonatto, Bica, \& Pavani \cite{BBP2004}). 

This paper is organized as follows. In Sect.~\ref{m16} we condense previous results on NGC\,6611. 
In Sect.~\ref{2mass} we present the 2MASS data, correct the photometry for differential reddening,
subtract the field-star contamination and analyze the CMDs and colour-colour diagrams. In Sect.~\ref{struc} 
we discuss the cluster structure. In Sect.~\ref{MF} we derive LFs and MFs and discuss stellar content 
properties. In Sect.~\ref{comp} we compare NGC\,6611 with nearby, older open clusters and discuss dynamical 
states. Finally, concluding remarks are given in Sect.~\ref{Conclu}. 

\section{The young open cluster NGC\,6611}
\label{m16}

The equatorial young open cluster NGC\,6611 is included in the Ser\,OB\,1 association and the W\,37 
molecular cloud. It is part of a larger sequence of star-forming events which have just begun in 
the region and are still embedded in the molecular cloud (Hillenbrand et al. \cite{Hillenbrand93}).
The cluster contains 8 O\,V-stars (Bosch, Morrell, \& Niemel\"a \cite{Bosch99}).  

The first comprehensive photometric study on this object was carried out by Walker (\cite{Walker61}) 
who obtained UBV photoelectric and photographic observations of NGC\,6611 limited at V=16.7.
He found that in the southern part of the region the colour excess is relatively small, 
$\ebv=0.6$, while to the north the reddening increases irregularly reaching $\ebv=1.73$. Stars 
above the MS were considered possible members in the gravitational contraction phase. 
He derived an age of $\rm 1.8\,Myr$ based on the analysis of 532 probable member stars.

Sagar \& Joshi (\cite{Sagar79}) derived a distance to the Sun $\rm\ds=3.2\pm0.3\,kpc$,
an age in the range 3.0 - 5.5\,Myr, and observed an extremely flat luminosity function in
this open cluster.

Based on photoelectric UBV photometry, Sagar et al. (\cite{Sagar86}) derive a flat MF 
with a slope $\chi\approx-0.2$ for stars in the mass range $\rm9\leq m(\ms)\leq70$.

By means of photometric observations over the range from 0.3 to 3.7\,$\mu$m Chini \& Wargau 
(\cite{cw90}) provide evidence of an abnormal extinction law only at wavelengths shorter than 
0.55\,$\mu$m. They conclude that all PMS stars previously identified by Walker (\cite{Walker61}) 
present reddening due to normal dust, therefore suggesting that they are foreground objects.

Hillenbrand et al. (\cite{Hillenbrand93}) combined CCD UBV observations with near-infrared JHK 
ones to establish a theoretical HR diagram for the cluster, from which they conclude that NGC\,6611
is actively forming 3 - 8\,\ms\ stars. Using the optical data they infer from a spatial coverage of 
$\approx1\,600\arcmin^2$ an overall MS MF slope of $\rm\chi=0.1 - 0.3$ in the mass range $\rm5\leq 
m(\ms)\leq85$. It is worth noting that only 4 stars have masses between 40 and 85\,\ms, for which 
they derive an age of $2\pm1$\,Myr. They also present evidence that one 30\,\ms\ star has been 
formed $\sim6$\,Myr ago. They derive $\ds=2.0\pm0.1$\,kpc.

\begin{figure*}
\begin{minipage}[b]{0.50\linewidth}
\includegraphics[width=\textwidth]{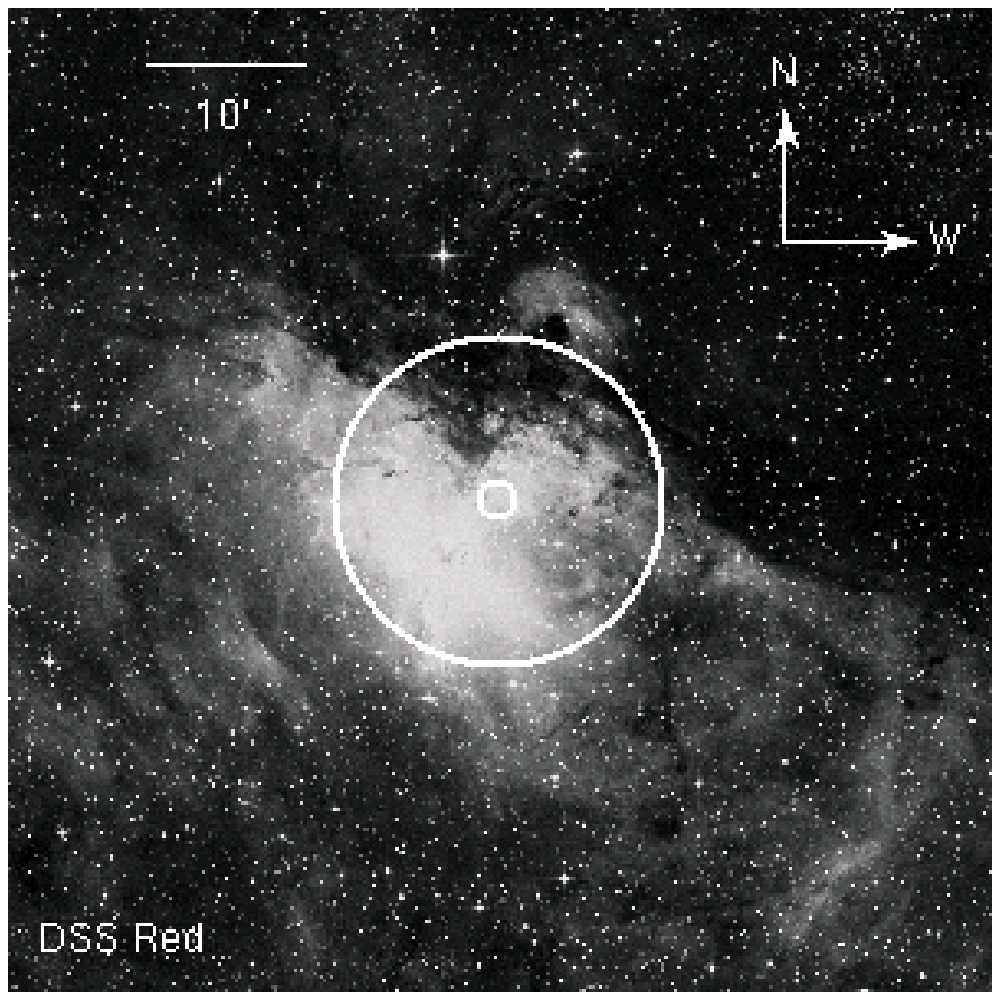}
\end{minipage}\hfill
\begin{minipage}[b]{0.50\linewidth}
\includegraphics[width=\textwidth]{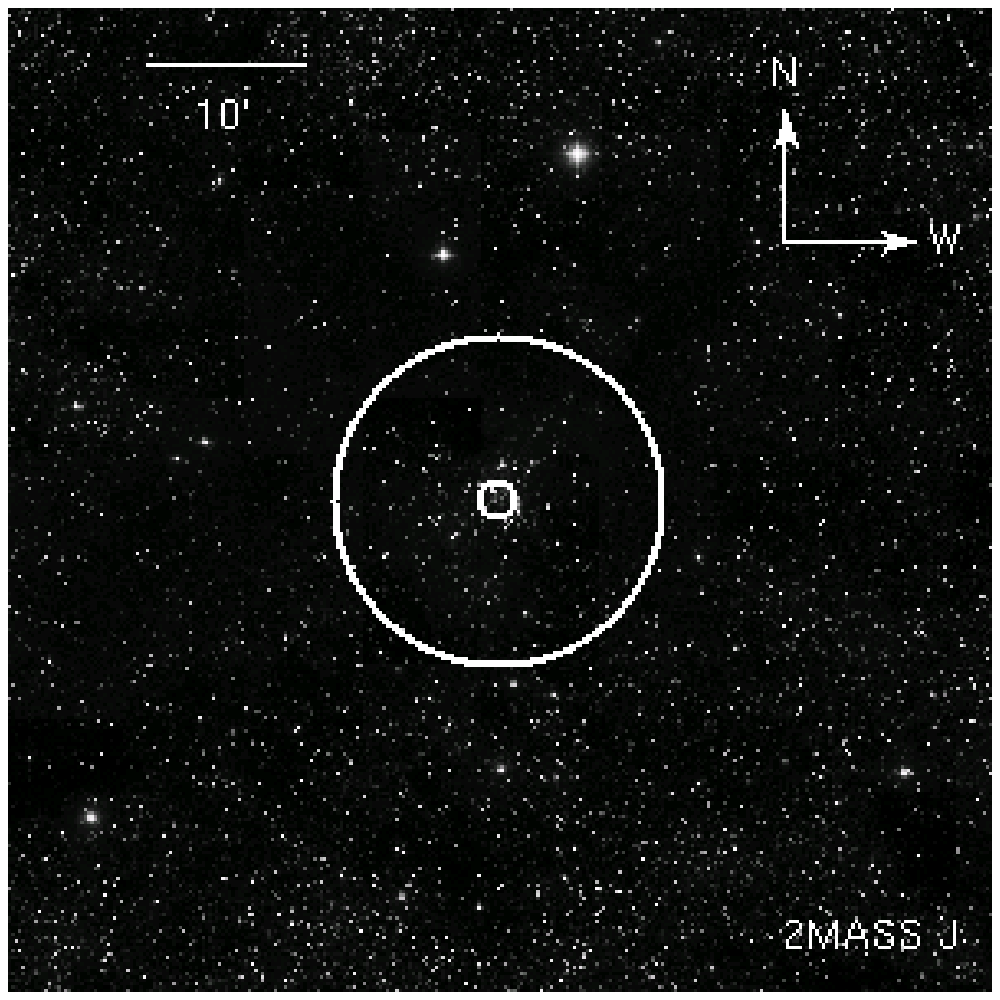}
\end{minipage}\hfill
\caption[]{Left panel: XDSS R image. Right panel: 2MASS J image. The area covered in
both images is $1^\circ\times1^\circ$. The small circle at the center of each image 
indicates the core, and the large one indicates the size of NGC\,6611 (limiting 
radius).}
\label{fig1}
\end{figure*}

By constructing new HR diagrams from the data published by Hillenbrand et al. (\cite{Hillenbrand93})
and applying a different mass calibration,  Massey et al. (\cite{Massey95}) derived an age in the
range 1-5\,Myr, $\chi=-0.3\pm0.2$, and a stellar density of $\rm\rho=5.4\times10^4\,kpc^{-2}$ for
stars above 10\,M$_{\odot}$.

Belikov et al. (\cite{Belikov99}) analyzed the spatial and proper motion distribution of cluster 
stars, assuming a normal distribution for the spatial velocities. They derived a distance to the 
Sun $\rm\ds=2.14\pm0.10\,kpc$ and obtained spatial dispersions of $\rm\sigma_{XY}^{core}=1.20\arcmin$ 
and $\rm\sigma_{XY}^{corona}=4.77\arcmin$, respectively for the core and corona subsystems in 
NGC\,6611. They used the same data to measure the differential reddening throughout the cluster
area.

Belikov et al. (\cite{Belikov00}) studied NGC\,6611 based on a compilation of photographic, 
photoelectric and CCD optical (UBV) observations covering a circular area of 45.6\arcmin\ in
diameter centered on NGC\,6611. They present radial profiles of the star density distribution in 
different azimuthal sectors. Small differences found between the density of the north-west and 
south-east sectors were attributed to irregularities of the absorption due to interstellar dust 
rather than to stellar density fluctuations. The south-east sector of the cluster is less affected 
by dust than the north-west one. The total radial density distribution allowed them to conclude that 
virtually no cluster members are present beyond the edge of the cluster corona, at 
$\rm 3\sigma_{XY}^{corona}=14.3\arcmin$. They derive an age of 6\,Myr for NGC\,6611. Based on the 
number frequency of member stars in the range $\rm 2.1\leq m(\ms)\leq85$ they infer a flat MF slope, 
$\chi=0.2$.

Tadross et al. (\cite{Tad2002}) presented an analysis of UBV CCD observations intending to derive 
morphological parameters for a large sample of open clusters. For NGC\,6611 they obtained a
colour excess $\ebv=0.68$, $\ds\approx1.6$\,kpc, age $\approx3.2$\,Myr, linear diameter 
$\rm D=3.6$\,pc, Galactocentric distance $\dgc=6.94$\,kpc, number of member stars 
$\rm N_*=282$, and a total mass of $\mtot=435\,\ms$.

The presence of significant numbers of PMS stars in NGC\,6611 has been confirmed by means of the 
observation of a large number of emission-line stars and IR disk signatures (Hillenbrand et al. 
\cite{Hillenbrand93}; Massey et al. \cite{Massey95}), which is an expected population in such a 
young cluster. Recently Oliveira et al. (\cite{Oliveira05}) obtained IZJHKL' photometry aiming 
at identifying PMS stars in NGC\,6611. Their Fig.~2 shows the CMDs $\rm I\times (I-Z)$ for the 
central area ($\rm7\arcmin\times6\arcmin$) and for a control field 16\arcmin\ away from the cluster 
center. The PMS population, defined within $\rm 12<I<19$, corresponding to the mass range $\rm 
0.4<m(\ms)<7$, stands out in the central area and decreases significantly towards the control field. 
The MS population does not show such trend, its spatial distribution remaining apparently constant 
throughout the fields.
   
In the left panel of Fig.~\ref{fig1} we provide an R XDSS image encompassing an area of 
$1^\circ\times1^\circ$ centered on NGC\,6611, extracted from the Canadian Astronomy Data Centre 
(CADC\footnote{\em http://cadcwww.dao.nrc.ca/}). In the right panel we show the corresponding 
2MASS J image. The stark contrast between the R and J images reflects primarily the effect of
the $\rm H\alpha$ emission in the R band and how dust is much more penetrable in the near-infrared 
bands. Thus,
the analysis of NGC\,6611 and its outskirts with near-infrared data is naturally justified  
because it allows access to low-mass stars with moderate aperture telescopes. In this sense,
Hillenbrand et al (\cite{Hillenbrand93}) investigated NGC\,6611 by means of deep near-infrared 
observations covering a $15\arcmin\times15\arcmin$ area with useful limits of J=16, H=14.5 and  
K=13.5, similar to those of 2MASS.

\begin{figure} 
\resizebox{\hsize}{!}{\includegraphics{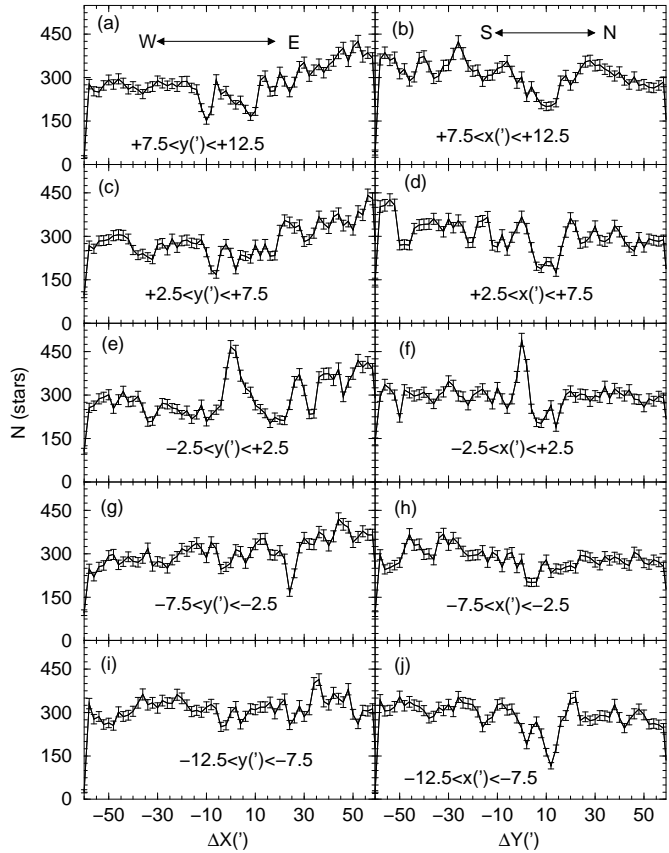}}
\caption[]{Spatial distribution of the number of stars (brighter than $\rm\jj=15.8$) around the 
central position of NGC\,6611. The density of stars increases to the East and South. $1\sigma$ 
Poisson errors are shown.}
\label{fig2}
\end{figure}

According to the WEBDA\footnote{\em http://obswww.unige.ch/webda} open cluster database 
(Mermilliod \cite{Merm1996}), the central coordinates of NGC\,6611 are (J2000) $\alpha=18^h18^m48^s$, 
and $\delta=-13^\circ48\arcmin24\arcsec$. However, the corresponding radial density profile 
(Sect.~\ref{struc}) presented a dip for $\rm R=0\arcmin$. Accordingly, we searched for a new center 
by examining histograms for the number of stars in 1\arcmin\ bins of right ascension and declination. 
The resulting coordinates which maximize the density of stars at the center are (J2000) $\alpha=18^h18^m40.8^s$, 
and $\delta=-13^\circ47\arcmin24.0\arcsec$, corresponding to $\ell=16.95^\circ$ and $b=0.83^\circ$. 
In what follows we refer to these optimized coordinates as the center of NGC\,6611. WEBDA gives a 
colour excess of $\ebv=0.768$, distance to the Sun $\ds=1.75$\,kpc, and $\rm age\approx7.6\,Myr$.

\section{Photometric parameters from 2MASS data}
\label{2mass}

The VizieR\footnote{\em http://vizier.u-strasbg.fr/viz-bin/VizieR?-source=II/246} tool was 
used to extract \jj, \hh\ and \ks\ 2MASS photometry of the stars present in a circular area 
with radius $\rm R=60\arcmin$ centered on the optimized coordinates described in the previous
section. The faint-magnitude limit of the extracted stars is brigther than that corresponding
to the 99.9\% Point Source Catalogue Completeness Limit\footnote{According to the Level\,1 
Requirement, according to 
{\em\tiny http://www.ipac.caltech.edu/2mass/releases/allsky/doc/sec6\_5a1.html }}, $\jj=15.8$, 
$\hh=15.1$\ and $\ks=14.3$, respectively.

As a first step in the investigation of NGC\,6611 we use the 2MASS data to examine in Fig.~\ref{fig2} 
the distribution of stars in the area surrounding the cluster. Each panel depicts the number of stars 
brighter than $\rm\jj=15.8$ in strips of 5\arcmin\ width in declination (left panels) and right 
ascension (right panels) with respect to the optimized cluster center. The rather concentrated structure 
of NGC\,6611 is apparent in panels (e) and (f). Within uncertainties the cluster appears to be spatially 
asymmetric, being more extended in right ascension (apparent diameter $\rm D_\alpha\sim21\arcmin$, 
(panel (e)) than in declination ($\rm D_\delta\sim12\arcmin$, (panel (f)). The spatial distribution of 
stars around the cluster is non-uniform, increasing significantly above the $1\sigma$ Poisson errors 
towards the East (panels (a), (c), (e) and (g)) and less markedly to the South (panels (b) and (d)). 
Part of this non-uniformity can be accounted for by differential reddening, which is important 
throughout the cluster area (Belikov et al. \cite{Belikov99}).

\begin{figure} 
\resizebox{\hsize}{!}{\includegraphics{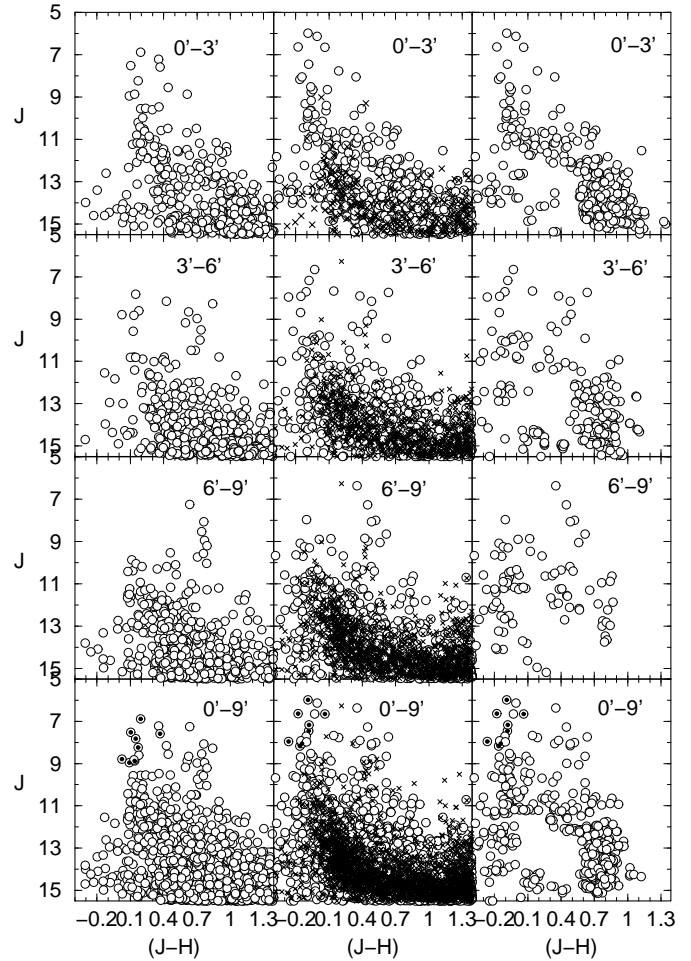}}
\caption[]{$\jj\times\jh$ CMDs of the central 9\arcmin\ of NGC\,6611. Left panels: observed 
photometry. Middle panels: reddening-corrected CMDs according to Belikov et al. (\cite{Belikov99}).
Right panels: reddening-corrected and field-star decontaminated CMDs. Note the extended MS and 
conspicuous PMS population, 
which is restricted essentially to the central 6\arcmin. Field stars are shown as x in the middle 
panels. Filled circles in the bottom panels: O\,V stars from Bosch, Morrell \& Niemel\"a 
(\cite{Bosch99}).}
\label{fig3}
\end{figure}

\subsection{Spatial variation of reddening}
\label{SVR}

Because of the low Galactic latitude ($b=0.83^\circ$), the field of NGC\,6611 presents significant 
field-star contamination. This can be seen in the left panels of Fig.~\ref{fig3} where we show the 
observed $\jj\times\jh$ CMDs of three consecutive circular regions ($\rm\Delta R=3\arcmin$) 
centered on NGC\,6611. Besides conspicuous MS stars, the integrated CMD up to $\rm R=9\arcmin$ (bottom
panel) presents important field-star contamination. 

As discussed in Belikov et al. (\cite{Belikov99}), the spatial field of NGC\,6611 is critically affected 
by differential reddening with important variations both in \ebv\ and $\rm R_V$ in areas separated 
by distances as small as $\sim2\arcmin$. They present \ebv\ and $\rm R_V$ values derived from spectroscopy 
and proper motion of cluster stars in $\rm2\arcmin\times2\arcmin$ cells covering an area of about 
$\rm40\arcmin\times36\arcmin$ centered on NGC\,6611. Note that \ebv\ in each cell corresponds to 
the absolute reddening value. 

We applied the above procedure to produce reddening-corrected 2MASS photometry for all stars in each 
cell to yield a total region reaching $\rm R\approx23\arcmin$ from the center of NGC\,6611. In the 
vacant cells we followed their suggestion and adopted the average values of \ebv\ and $\rm R_V$.  For 
reddening and absorption transformations in the 2MASS bands we use the relations $\rm A_J=0.276\times 
R_V\,\ebv$, $\rm A_H=A_J-0.33\,\ebv$ and $\rm A_{K_S}=A_J- 0.49\,\ebv$, according to Dutra, Santiago 
\& Bica (\cite{DSB2002}, and references therein).

The resulting reddening-corrected CMDs are shown in the middle panels of Fig.~\ref{fig3}. Comparison
of the reddening-corrected CMDs with the observed ones shows that the average colour-excess throughout 
the cluster area is $\langle\ejh\rangle=0.25$, corresponding to $\langle\ebv\rangle=0.8$.

\subsection{Field-star decontamination}
\label{FSD}

Finally, to uncover the intrinsic CMD morphology we apply a field-star decontamination procedure. 
Because of the spatial limitation on the useful area caused by the differential-reddening correction 
(Sect.~\ref{SVR}), we consider as offset field the ring located at $\rm17\arcmin\leq R\leq22\arcmin$.
This area is large enough to produce statistical representativity of the field stars, both in magnitude 
and colours.

To illustrate the amount and colour-magnitude distribution of the contamination in the area of NGC\,6611 
we plot in the middle panels of Fig.~\ref{fig3} the corresponding (same area) field-star contributions
taken from rings with external radius at 22\arcmin. As a consequence we can see that most of the faint 
stars - in particular those mimicking a low-MS - are in fact field-stars.

Based on the spatial number-density of stars in the offset field, the decontamination procedure estimates 
the number of field stars which within the $\rm1\sigma$ Poisson fluctuation should be present in the 
cluster field. The observed CMD is then divided in colour/magnitude cells from which stars are randomly 
subtracted in a number consistent with the expected number of field stars in the same cell. The 
dimensions of the colour/magnitude cells can be subsequently changed so that the total number of stars 
subtracted throughout the wole cluster area matches the expected one, within the $\rm1\sigma$ Poisson 
fluctuation. Since the field stars are taken from an external annulus of fixed dimensions, corrections 
are made for the different solid angles of cluster and offset field. This procedure can be applied
to the cluster region as a whole or internal regions as well. Note that because this procedure actually
excludes stars from the original files - thus artificially changing both the radial distribution of stars
and the LF - we use it only for the sake of uncovering the intrinsic CMD morphology. 

The resulting field-star decontaminated (differential-reddening corrected) CMDs are shown in the right 
panels of Fig.~\ref{fig3}. As expected, most of the faint-star distribution was subtracted by the
decontamination procedure. The central region ($\rm0\arcmin\leq R\leq3\arcmin$) presents an extended,
vertical MS reaching from $\jj\approx11.5$ to $\jj\approx5.5$ which reflects the young age of NGC\,6611. 
This region also presents a conspicuous population of PMS stars ($\rm0.54\leq\jh\leq1.1$) which extends 
essentially up to $\rm R\approx6\arcmin$. The entry point (turn-on stars) of the PMS into the MS occurs 
at $\jj\approx11.1$. The region $\rm6\arcmin\leq R\leq9\arcmin$\ is sparsely populated by cluster stars, 
with a small fraction of MS and PMS stars still discernible.

\begin{figure} 
\resizebox{\hsize}{!}{\includegraphics{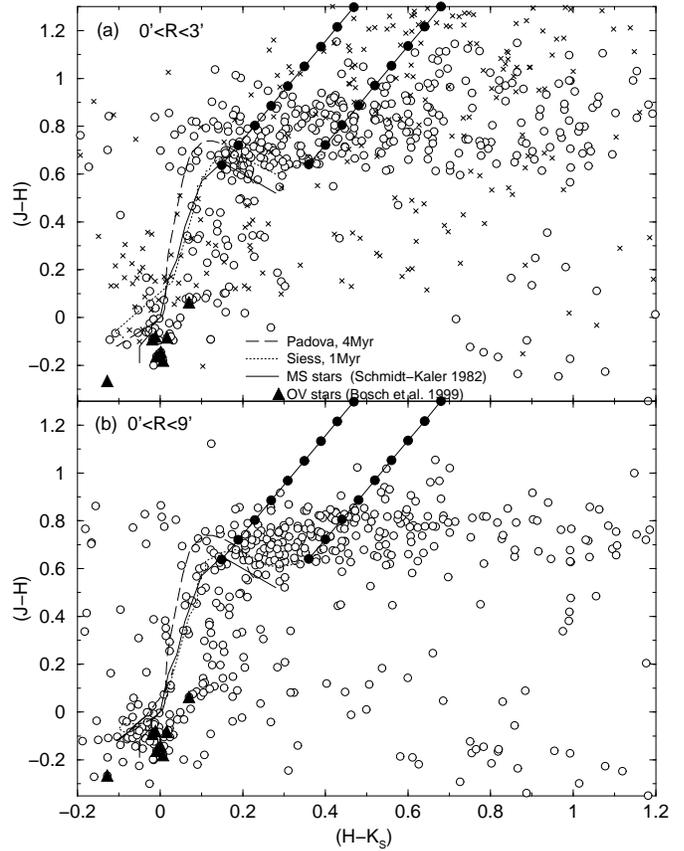}}
\caption[]{$\rm\hk\times\jh$ diagram of the central 3\arcmin\ (Panel (a)) and 9\arcmin\ (Panel (b)) 
of NGC\,6611. Dashed line: 4\,Myr Padova isochrone. Dotted line: 1\,Myr sequence of PMS stars 
(Siess, Dufour, \& Forestini \cite{Siess2000}).
Solid line: MS stars from Schmidt-Kaler (\cite{SK82}). Filled triangles: O\,V stars (Bosch, Morrell, 
\& Niemel\"a \cite{Bosch99}). Filled circles: reddening vectors; $\rm\Delta\ebv=0.25$ between 2
consecutive circles. The corresponding field stars are shown in Panel (a) as `x'. Both diagrams were 
produced after field-star decontamination.} 
\label{fig4}
\end{figure}

\subsection{Age}
\label{age}

Deriving a precise age for clusters younger than $\sim10$\,Myr by means of isochrone fit is a 
difficult task because of the nearly-vertical MS and lack of post-MS stars. As a consequence of
the scarcity of observational constraints in the CMD, several age-solutions are possible. In the 
case of NGC\,6611, the CMD (Fig.~\ref{fig3}) can be acceptably fitted with isochrones with ages up 
to 10\,Myr. In the present case we use the set of solar-metallicity isochrones of Padova (Girardi 
et al. \cite{Girardi2002}) computed with the 2MASS \jj, \hh\ and \ks\ filters\footnote{\em\tiny 
http://pleiadi.pd.astro.it/isoc\_photsys.01/isoc\_photsys.01.html}. The 2MASS transmission filters 
produced isochrones very similar to the Johnson ones, with differences of at most 0.01 in \jh\ 
(Bonatto, Bica \& Girardi \cite{BBG2004}).

On the other hand, properties of young/embedded objects, in particular the age, can be 
alternatively inferred by means of infrared colour-colour diagrams (e.g. Lada, Alves \& Lada 
\cite{LAL96}). The fraction of stars with \ks\ excess in embedded clusters is an age indicator 
(e.g. Lada, Alves \& Lada \cite{LAL96}; Soares \& Bica \cite{SB02}) in the sense that larger fractions 
correspond to younger evolutionary stages. In Fig.~\ref{fig4} we show the $\rm\hk\times\jh$ diagram
of the central 3\arcmin\ (Panel (a)) and 9\arcmin\ (Panel (b)) of NGC\,6611. Both diagrams were
produced after differential-reddening correction and field-star decontamination. To compare stars 
in different evolutionary stages we show in Panel (a) the loci occupied by MS stars according to the 
youngest Padova isochrone (4\,Myr) and to those of Schmidt-Kaler (\cite{SK82}). The 1\,Myr sequence 
of PMS stars (Siess, Dufour, \& Forestini \cite{Siess2000}), and the loci of the O\,V stars with
available spectroscopic data of Bosch, Morrell, \& Niemel\"a (\cite{Bosch99}) are shown as well. 
Extinction effects can be estimated by the reddening vectors.

For NGC\,6611 we measured within $\rm R\leq3\arcmin$ a fraction of $\rm f_{exc}=0.54\pm0.05$ of
\ks\ excess stars. Considering uncertainties this value is intermediate between those of the
$\rho$ Ophiuchi cluster ($\rm f_{exc}=0.5 - 0.7$ -- Greene \& Young \cite{GY92}; Strom, Kepner,
\& Strom \cite{Strom95}) with an age of 1\,Myr or less (Greene \& Meyer \cite{GM95}) and that of 
the Taurus Dark Nebula group ($\rm f_{exc}=0.5$ -- Kenyon \& Hartmann \cite{KH95}) with an age of 
1.5\,Myr. Based on this we estimate an age of $\rm1.3\pm0.3$\,Myr for NGC\,6611. These values are 
similar to those of Trapezium, NGC\,2327 cluster and BRC\,27 cluster (Soares \& Bica \cite{SB02}). 

This estimate is consistent with {\it (i)} the presence of a significant population 
of PMS (e.g. Walker \cite{Walker61}; Hillenbrand et al. \cite{Hillenbrand93}; Massey et al. 
\cite{Massey95}; Oliveira et al. \cite{Oliveira05}), {\it (ii)} the fraction of $\sim58\%$ of 
PMS having circumstellar discs (e.g. Hillenbrand et al. \cite{Hillenbrand93}; Massey et al. 
\cite{Massey95}; Oliveira et al. \cite{Oliveira05}), and {\it (iii)} the fact that NGC\,6611 
is ionizing the \ion{H}{ii} region Sh2-49 (e.g. Bosch, Morrell, \& Niemel\"a \cite{Bosch99}; 
Hillenbrand et al. \cite{Hillenbrand93}).

\subsection{Distance from the Sun}
\label{dsun}

\begin{figure} 
\resizebox{\hsize}{!}{\includegraphics{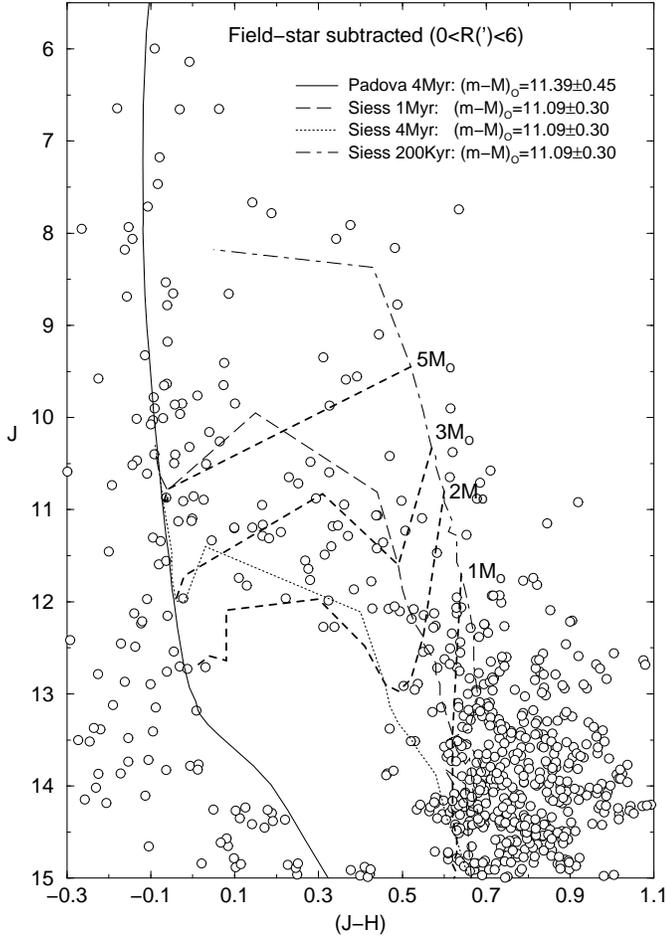}}
\caption[]{The 4\,Myr Padova isochrone set to the reddening-corrected, field-star decontaminated 
CMD of the central 6\arcmin\ of NGC\,6611 according to $\ds=1.92\pm0.45$\,kpc. PMS model isochrones 
of Siess, Dufour, \& Forestini (\cite{Siess2000}) with ages of 200\,Kyr, 1\,Myr and 4\,Myr are fitted 
to the PMS population resulting in $\rm\ds=1.65\pm0.27\,kpc$. Evolutionary tracks for PMS stars of 
masses 1, 2, 3and 5\,\ms\ are shown as dotted lines. }
\label{fig5}
\end{figure}

Similarly to the age, the lack of evolved features in the CMD of NGC\,6611 precludes a precise
distance determination based directly on isochrone fit. In this sense we will derive the distance 
of NGC\,6611 from the Sun using two independent methods. Bosch, Morrell \& Niemel\"a (\cite{Bosch99}) 
derived spectroscopic data of 8 MS O member stars (which are identified in the bottom panels of 
Fig.~\ref{fig3}). Each O\,V star was corrected for reddening according to the corresponding 
cell \ebv\ and $\rm R_V$ values in Belikov et al. (\cite{Belikov99}). From these stars we derive 
an absolute average distance modulus of $\rm\mMo=11.39\pm0.46$, corresponding to a distance from 
the Sun $\rm\ds=1.92\pm0.45\,kpc$. We use this value to set the 4\,Myr Padova isochrone in the 
reddening-corrected, field-star decontaminated CMD of the central 6\arcmin\ of NGC\,6611 in 
Fig.~\ref{fig5}. Within uncertainties, the least-massive stars already in the MS of NGC\,6611 have 
mass in the range 3--5\,\ms.

Alternatively, we estimate the distance of NGC\,6611 by means of the MS entry point of the PMS 
stars. To do this we use the turn-on locus in the reddening-corrected, field-star decontaminated 
CMD of the central 6\arcmin\ in Fig.~\ref{fig5}. The PMS stars in Fig.~\ref{fig5} turn out to be 
basically contained within the 200\,Kyr and 4\,Myr isochrones (Siess, Dufour, \& Forestini 
\cite{Siess2000}), after applying an absolute distance modulus of $\rm\mMo=11.09\pm0.30$.
This corresponds to a distance from the Sun of $\rm\ds=1.65\pm0.27\,kpc$ which, within the
uncertainties, agrees with the previous value based on the O\,V stars. For completeness we also 
show in Fig.~\ref{fig5} evolutionary tracks up to the zero-age main sequence (ZAMS) for PMS stars 
with masses 1, 2, 3 and 5\,\ms. Similarly to the value based on the 4\,Myr Padova isochrone, the 
low-mass MS of NGC\,6611 implied by the PMS occurs in the range 3--5\,\ms.

Based on the above estimates we adopt the average value $\rm\ds=1.8\pm0.5\,kpc$ as the distance of 
NGC\,6611 from the Sun. With this value the Galactocentric distance of NGC\,6611 turns out to be 
$\dgc=6.3\pm0.5$\,kpc, using 8.0\,kpc as the distance of the Sun to the center of the Galaxy (Reid 
\cite{Reid93}). The present value of the distance from the Sun agrees with that derived by Tadross 
et al. (\cite{Tad2002}) and is comparable to that of WEBDA (1.75\,kpc), but falls $\sim0.5$\,kpc 
short with respect to the average value found in the literature. 

\section{Cluster structure}
\label{struc}

\begin{figure} 
\resizebox{\hsize}{!}{\includegraphics{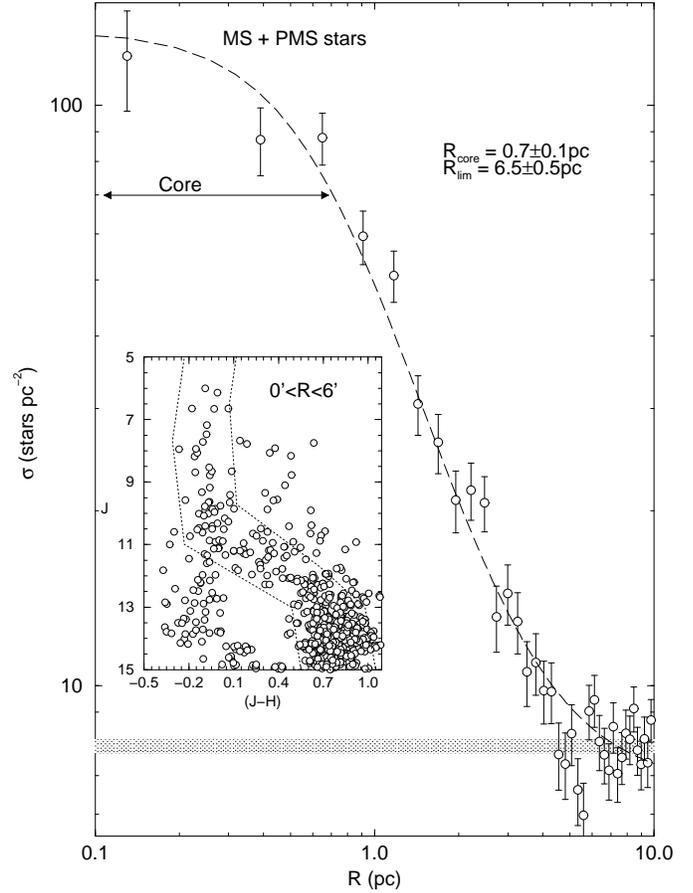}}
\caption[]{Radial density profile of the cluster NGC\,6611. Dashed line: 2-parameter King profile. Shaded 
region: stellar background level (average number of stars in the ring $17\arcmin\leq R\leq22\arcmin$). 
The dimension of the core is indicated. The colour-magnitude filter used to discard stars with discrepant 
colours is shown in the inset, superimposed on the reddening-corrected and field-star decontaminated CMD 
of the central 6\arcmin.}
\label{fig6}
\end{figure}

Structural parameters of NGC\,6611 were derived by means of the radial density profile, defined as the 
projected number of stars per area around the cluster center. In the case of this low-latitude cluster
the contamination of the CMD by field stars must be taken into account in order to derive the intrinsic
radial distribution of stars. We do this by considering the colour-magnitude filter shown in the inset 
of Fig.~\ref{fig6} which, within uncertainties, describes the cluster CMD morphology from the upper MS 
to the PMS stars. Before counting stars we applied the colour-magnitude filter to the differential-reddening 
corrected CMD of the cluster (including stars with distance to the center from 0\arcmin\ to 23\arcmin), to 
discard stars with discrepant colours. The colour-magnitude filtering procedure has been previously 
applied in the analysis of the open clusters M\,67 (Bonatto \& Bica \cite{BB2003}), NGC\,188 (Bonatto, 
Bica \& Santos Jr. \cite{BBS2004}) and NGC\,3680 (Bonatto, Bica \& Pavani \cite{BBP2004}). The radial 
density profile was obtained by counting stars inside concentric annuli with a step of 0.5\arcmin\ 
in radius. In the present case the background contribution level corresponds to the average number 
of stars included in the ring located at $\rm 17\arcmin\leq R\leq 22\arcmin$ ($\rm 8.8\leq 
R(pc)\leq11.4$), resulting in $\rm\sigma_{bg}=2.14\pm0.06\,stars\,(\arcmin)^{-2}=
7.8\pm0.2\,stars\,pc^{-2}$. Note that the colour-magnitude filter encompasses both the MS 
and PMS stars.

The resulting radial density profile of the $\rm MS + PMS$ stars is shown 
in Fig.~\ref{fig6}. For absolute comparison between clusters we scale the radius in the abscissa 
in parsecs, and the number density of stars in the ordinate in $\rm stars\,pc^{-2}$\ using the 
distance derived in Sect.~\ref{2mass}. The statistical significance of the profile is reflected 
in the $1\sigma$\ Poisson error bars. 

Structural parameters of NGC\,6611 were derived by fitting the two-parameter King (\cite{King1966a}) 
surface density profile to the background-subtracted radial distribution of stars. The two-parameter 
King model essentially describes the intermediate and central regions of normal clusters (King 
\cite{King1966b}; Trager, King \& Djorgovski \cite{TKD95}). The fit was performed using a nonlinear 
least-squares fit routine which uses the errors as weights. The best-fit solution is shown in 
Fig.~\ref{fig6} superimposed on the observed radial density profile. Parameters derived are the King 
background-subtracted central density of stars $\rm\sigma_{0K}=35\pm6\,stars\,(\arcmin)^{-2}=128\pm20
\,stars\,pc^{-2}$, and the core radius $\rm\rc=1.35\pm0.15\arcmin=0.7\pm0.1\,pc$. Considering the 
radial density profile fluctuations with respect to the background level, we can 
define a limiting radius (\rl) for the cluster, in the sense that for regions beyond $\rl$, the 
null-contrast between cluster and background star density would produce large Poisson 
errors and consequently, meaningless results. Thus, for practical purposes, the 
bulk of the cluster stars are contained within $\rl$. For NGC\,6611 we estimate a limiting radius
$\rm\rl=12.5\pm1.0\arcmin=6.5\pm0.5\,pc$.

Despite the young age of NGC\,6611 the King profile provides a good fit of the  stellar content 
in this cluster, within uncertainties. Since it follows from an isothermal (virialized) sphere, 
the close similarity of the radial distribution of stars in NGC\,6611 with a King profile may 
suggest that the internal structure of this cluster (particularly the core) has reached some 
level of energy equipartion. However, part of this effect appears to be related to formation 
processes as well (Sects.~\ref{dyna} and \ref{comp}). 

\section{Luminosity and mass functions of the MS}
\label{MF}

The rather populous nature of NGC\,6611 provides an opportunity to study the spatial distribution 
of LFs and MFs $\left(\phi(m)=\frac{dN}{dm}\right)$ of the MS stars in such a young open cluster. 

Based on the King profile fit of the MS radial density profile of NGC\,6611 (Fig.~\ref{fig6}) we 
decided to study the MF in the following regions: {\em (i)} $\rm 0.0\leq R(pc)\leq0.7$ (core), 
{\em (ii)} $\rm 0.7\leq R(pc)\leq6.5$ (halo), and {\em (iii)} $\rm 0.0\leq R(pc)\leq6.5$\ (overall). 
In order to maximize the significance of background counts, we consider as offset field the 
outermost ring at $\rm 8.8\leq R(pc)\leq 11.4$, which lies $\sim2.3$\,pc beyond the limiting 
radius. 

\begin{figure} 
\resizebox{\hsize}{!}{\includegraphics{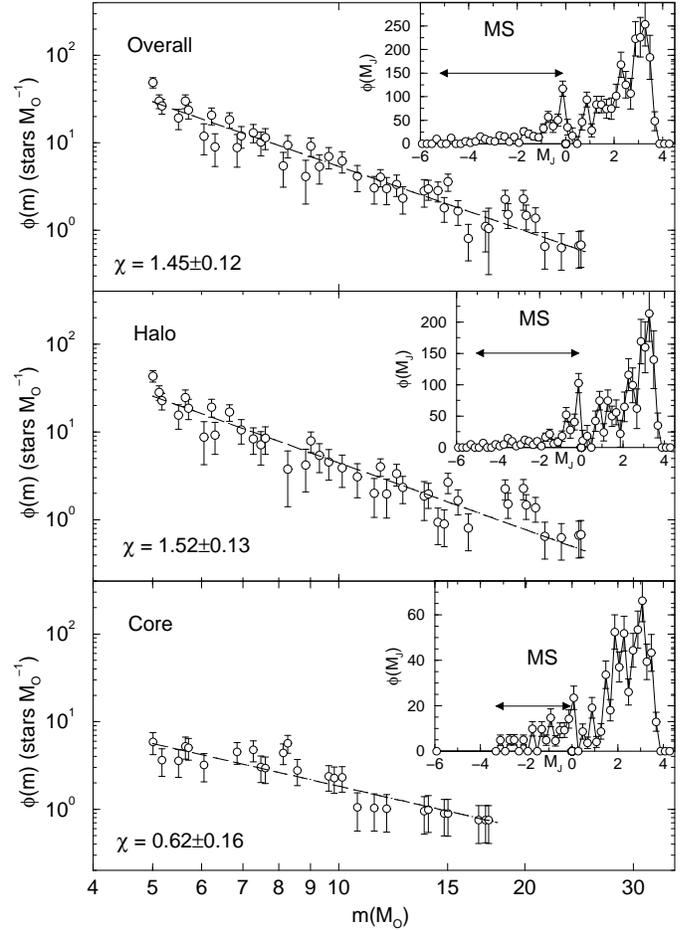}}
\caption[]{Mass functions of the MS stars (empty circles) in different spatial regions of 
NGC\,6611. Each panel contains MFs derived from the \jj, \hh\ and \ks\ 2MASS photometry. MF 
fits $\left(\phi(m)\propto m^{-(1+\chi)}\right)$ are shown as dashed lines, and the respective MF 
slopes are given. The insets show the $\rm MS+PMS$ field-star subtracted LFs in each spatial region.}
\label{fig7}
\end{figure}

Similarly to the structural analysis (Sect.~\ref{struc}) the first step in the present analysis
is to take into account the field-star contamination. In this case we apply the colour-magnitude 
filter (Fig.~\ref{fig6}) restricted to the MS stars (i.e. $\jj\leq11.4$). The filtering process 
takes into account most of the field stars, leaving only a residual contamination. We deal with 
this residual contamination statistically by building LFs for each cluster region and offset field 
($17\arcmin\leq R\leq22\arcmin$) separately. We consider the three 2MASS bands independently when
building the LFs. Note that the faint-magnitude limit of the MS is significantly brighter than that 
of the 99.9\% Point Source Catalogue Completeness Limit (Sect.~\ref{2mass}). We take $\rm\jj=5$ as 
the upper-MS limit (Fig.~\ref{fig3}). We build the LF of each 2MASS band by counting stars in magnitude 
bins from the respective faint magnitude limit to the upper MS, both for each cluster region and offset 
field. Considering that the solid angle of the offset field may be different from that of a given 
cluster region, we multiply the offset field LF by a numerical factor so that the solid angles match. 
The intrinsic LF of each cluster region is obtained by subtracting the respective (i.e. solid angle-corrected) 
offset field LF from that of the cluster region. Finally, the intrinsic LFs are transformed into MFs 
using the mass-luminosity relation obtained from the 4\,Myr Padova isochrone (the youngest one available) 
and the absolute distance modulus $\mMo=11.39$ (Sect.~\ref{2mass}). These procedures are repeated 
independently for the three 2MASS bands. The final MF of a given cluster region is produced by 
combining the \jj, \hh\ and \ks\ MFs into a single MF. The resulting core, halo and overall MFs of 
NGC\,6611, covering the mass range $\rm5\leq m(\ms)\leq25$, are shown in Fig.~\ref{fig7}. 
To these MFs we fit the function $\phi(m)\propto m^{-(1+\chi)}$. The resulting fits are shown in 
Fig.~\ref{fig7}, and the MF slopes are given in col.~4 of Table~\ref{tab2}.

We show in each panel of Fig.~\ref{fig7} the corresponding field-star subtracted LFs, where the 
relative contribution of the PMS ($\rm M_J\geq-0.12$) with respect to the MS can be evaluated. 

Both the overall and halo MFs are slightly steeper than a typical Salpeter (\cite{Salpeter55}) IMF 
($\chi=1.35$); however, the core MF turned out to be flat, with $\chi=0.62\pm0.16$. The difference in 
MF slope between the core and halo ($\chi=1.52\pm0.13$, panel (b) of Fig.~\ref{fig7}), may be a 
consequence of mass segregation in the core. We discuss this point further in Sect.~\ref{dyna}. 

\subsection{Total observed mass}
\label{TM}

We provide in Table~\ref{tab2} parameters derived from the MFs of each spatial region considered. 
The number of PMS stars in the core, halo and overall regions was estimated upon integration 
of the respective field star-subtracted LFs (Fig.~\ref{fig7}) for $\rm M_J\geq-0.12$. 
Considering the observed distribution of PMS stars and the respective evolutionary tracks 
(Fig.~\ref{fig5}) and the fact that MFs in general increase in number for the subsolar-mass
range, for simplicity we assume an average PMS mass of 1\,\ms. The resulting number and mass of 
PMS stars are in cols.~2 and 3, respectively The observed number of MS stars and corresponding 
mass (cols.~5 and 6, 
respectively) are derived by integrating the MF through the mass range 5--25\,\ms. We add to 
these the corresponding values of the number and mass of PMS stars to derive the total number of 
observed stars (col.~7), observed mass (col.~8), projected mass density (col.~9) and mass density 
(col.~10).
 
\begin{table*}
\caption[]{Parameters derived from the MFs and PMS stars.}
\label{tab2}
\renewcommand{\tabcolsep}{0.7mm}
\renewcommand{\arraystretch}{1.6}
\begin{tabular}{cccccccccccccccc}
\hline\hline
&\multicolumn{2}{c}{PMS}&&\multicolumn{4}{c}{Observed MS}&
&\multicolumn{4}{c}{$\rm Observed\ MS+PMS$}&&$\tau$\\
\cline{2-3}\cline{5-8}\cline{10-13}\\
Region&N$^*$&m &&$\chi_{5-25}$&&$\rm N^*$&\mobs&&$\rm N^*$&\mtot&$\sigma$&$\rho$\\
(pc)&($10^2$stars)&($10^2\ms$)& &&&($10^2$stars)&($10^2\ms$)&& ($10^2$stars)&
($10^2\ms$)&($\rm \ms\,pc^{-2}$)&($\rm \ms\,pc^{-3}$)\\
 (1)& (2) & (3) && (4) &&(5) &(6) && (7)& (8) & (9) & (10)&&(11)\\
\hline
Core&$1.0\pm0.3$&$1.0\pm0.3$&&$0.62\pm0.16$& &$0.3\pm0.2$&$2.3\pm1.3$
&&$1.2\pm0.4$&$4.1\pm1.6$&$265\pm107$&$284\pm115$&&$1.8\pm0.6$\\

Halo&$2.6\pm0.8$&$2.6\pm0.8$ &&$1.52\pm0.13$& &$0.8\pm0.3$&$6.9\pm2.9$
 &&$3.4\pm0.9$&$9.6\pm3.1$&$7.3\pm2.3$&$0.83\pm0.27$&&---\\
 
Overall&$3.6\pm1.2$&$3.6\pm1.2$ &&$1.45\pm0.13$& &$0.9\pm0.3$&$8.4\pm3.2$ 
&&$4.5\pm1.2$&$12.0\pm3.4$&$9.0\pm2.6$&$1.0\pm0.3$&&$0.067\pm0.022$\\ 

\hline
Overall$^\dag$&$3.6\pm1.2$&$3.6\pm1.2$ &&$1.45\pm0.13$& &$1.0\pm0.3$&$12.7\pm3.5$ &&$4.6\pm1.3$&$16.3\pm3.5$&$12.3\pm2.6$&$1.4\pm0.3$&&$0.066\pm0.022$\\
\hline\hline
\end{tabular}
\begin{list}{Table Notes.}
\item Observed mass range: $\rm5-25\,\ms$. Col.~4 gives the MF slope derived for the MS stars. 
Col.~11: dynamical-evolution parameter $\rm\tau=age/t_{relax}$. ($\dag$): includes stars more 
massive than 25\,\ms.
\end{list}
\end{table*}

To complete the census of the mass stored in observed stars in NGC\,6611 we must include 
as well the stars more massive than 25\,\ms, which although present in the photometry, have not been 
counted in the mass estimate owing to the adopted MF upper cutoff ($\rm25\,\ms$). Hillenbrand et al. 
(\cite{Hillenbrand93}) give the number of stars in 
selected mass bins for massive stars: 5 stars with mass in the range $\rm 
25\leq m(\ms)\leq40$, 4 with $\rm 40\leq m(\ms)\leq60$, and 1 with $\rm 60\leq m(\ms)\leq85$. 
Accordingly, the mass stored in stars more massive than 25\,\ms\ amounts to $\approx435\,\ms$, 
about 50\% of the mass implied by the MF of the main sequence range considered above 
(col.~6 of Table~\ref{tab2}). For simplicity we add their mass to the overall region. To this 
effect we include an additional entry in Table~\ref{tab2} to account for the massive stars,
resulting in $\rm\mtot\sim1630\,\ms$. This value is $\sim4$\ times larger than the estimate of 
Tadross et al. (\cite{Tad2002}). The total observed mass of NGC\,6611 is a factor $\approx2$ larger than
the mass estimated for the Trapezium cluster (Lada \& Lada \cite{LL2003}). It should be noted
that the above mass and density estimates (cols.~(8) - (10) in Table~\ref{tab2}) are in fact lower limits,
since we are not taking into account stars less massive than 5\,\ms, a mass range for which we have no
information on the behaviour of the MF slope.

\subsection{Dynamical state of NGC\,6611}
\label{dyna}

The overall MF slope ($\chi=1.45\pm0.12$) in the mass range $\rm 5\leq m(\ms)\leq 25$ is similar
to that of a standard Salpeter ($\chi=1.35$) IMF. Besides, the MF slope presents large 
variations in the inner regions, being flat ($\chi=0.62\pm0.16$) in the core and rather steep 
($\chi=1.52\pm0.13$) in the halo (Table~\ref{tab2} and Fig.~\ref{fig7}). In older clusters
this fact reflects the effects of large-scale mass segregation, in the sense that low-mass stars 
originally in the core are transferred to the cluster's outskirts while massive stars accumulate 
in the core. This produces a flat MF in the core and a steep one in the halo (see, e.g. Bonatto
\& Bica \cite{BB2005}).

Mass segregation in a star cluster scales with the relaxation time, defined as $\rm 
\tr=\frac{N}{8\ln N}\tcr$, where $\rm\tcr=R/\sigma_v$ is the crossing time, N is the (total) number 
of stars and $\rm\sigma_v$\ is the velocity dispersion (Binney \& Tremaine \cite{BinTre1987}). 
The characteristic time scale in which a cluster reaches some level of kinetic energy equipartition
is described by $\tr$, when massive stars sink to the core and low-mass stars are transferred to 
the halo. Assuming a typical $\rm\sigma_v\approx3\,\kms$ (Binney \& Merrifield \cite{Binney1998}) 
we obtain for the overall cluster $\rm\tr\sim20\pm5$\,Myr, and for the core $\rm\tr\sim0.7\pm0.2$\,Myr.
Considering that NGC\,6611 is $\sim1.3$\,Myr old (Sect.~\ref{2mass}), the age of this cluster is $\sim2$ 
times larger than $\rm\tr(core)$. Thus, the presence of mass segregation and consequently some
degree of MF slope flattening can be expected in the core. However, the ratio cluster age to
\tr\ drops to $\sim0.07$ for the overall cluster, which is consistent with the typical Salpeter
slope and the absence of important mass segregation. 

Although OB stars in NGC\,6611 host more companions (for binaries with mass ratios $\rm q\geq0.1$) 
than solar-type field stars (Duch\^ene et al. \cite{Duch2001}), the expected influence of binaries 
in the MF would be to steepen the slope of a standard Salpeter IMF to $\chi\approx1.7$, and not to 
flatten it out (Sagar \& Richtler \cite{SagRich91}; Kroupa \cite{Kroupa2004}).

\section{Comparison with clusters in different dynamical-evolution states}
\label{comp}

In the previous sections we derived a series of parameters related to the structure and
dynamical evolution of NGC\,6611. At this point it may be useful to check how this young
cluster fits in the context of open clusters with different ages and in more advanced 
dynamical states. Bonatto \& Bica (\cite{BB2005}) presented a systematic analysis of a
set of open clusters with ages in the range 70 - 7\,000\,Myr and masses from 400 - 
5\,300\,\ms. The methodology used follows the same lines as that used the present paper.
As a result, a set of uniform parameters related to the structure (core and overall radii, mass
and density), dynamical state (core and overall MF slopes, evolutionary parameter $\rm\tau=age/\tr$),
as well as age and Galactocentric distance of open clusters was obtained. Some correlations 
among these parameters were verified, and a separation of massive ($m\geq1\,000\,\ms$) and
less-massive ($m\leq1\,000\,\ms$) clusters was observed. The core and overall dynamical-evolutionary 
parameter of NGC\,6611 is given in col.~11 of Table~\ref{tab2}.

\begin{figure} 
\resizebox{\hsize}{!}{\includegraphics{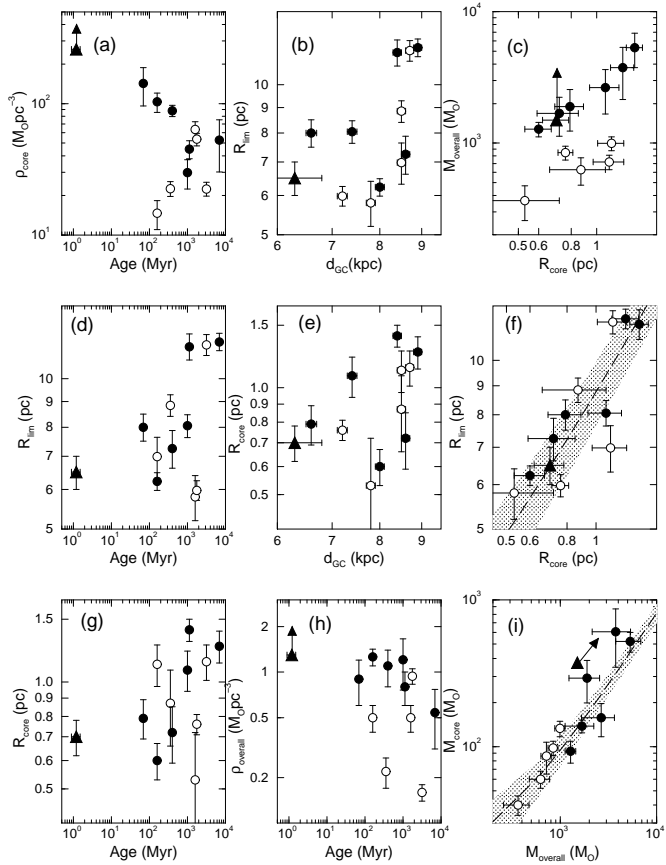}}
\caption[]{Relations involving structural parameters of open clusters. Filled circles:
clusters more massive than 1\,000\,\ms. Open circles, $\rm m<1\,000\,\ms$. Filled triangle:
NGC\,6611. Dashed lines: least-squares fits to the nearby clusters (see text). Shaded areas: 
$\rm1\sigma$ borders of the least-squares fits. Arrows indicate lower-limit estimates 
of mass and density for NGC\,6611.}
\label{fig8}
\end{figure}

In Fig.~\ref{fig8} we compare NGC\,6611 with the set of nearby, well-studied older clusters 
(Bonatto \& Bica \cite{BB2005}) in terms of structural parameters. Note that the 
mass and density values of NGC\,6611 are lower limits (Sect.~\ref{TM}).
The limiting radius of NGC\,6611 fits in the low-limit correlation of \rl\ with age 
(panel (d)). Although with a larger scatter, the same is observed in the correlation of \rc\
with age (panel (g)). The core density of NGC\,6611 follows the trend presented by massive 
clusters for young ages (panel (a)). A similar trend is seen for the overall density (panel (h)).
In Galactocentric distance the limiting radius of NGC\,6611 helps defining
a correlation (panel (b)) in the sense that clusters at larger \dgc\ tend to be larger, which
agrees with the results of Lyng\aa\ (\cite{Lynga82}) and Tadross et al. (\cite{Tad2002}).
The large scatter in panel (e) precludes any conclusion with respect to a dependence of the core
radius with \dgc. NGC\,6611 fits well in the tight correlations of core and overall mass (panel (i);
least-squares fit: $\rm M_{core}=(14.71\pm10.30)+(0.08\pm0.01)M_{overall}$ with a correlation 
coefficient $\rm CC=0.92$),
and core and limiting radii (panel (f); $\rm \rl=(1.05\pm0.45)+(7.73\pm0.66)\rc$ with   
$\rm CC=0.95$). Finally, in Bonatto, \& Bica (\cite{BB2005}) we observed
that massive and less-massive clusters follow parallel, different paths in the plane 
$\rm core\ radius \times overall\ mass$. NGC\,6611 consistently fits in the 
the massive clusters path (panel (c)). 

\begin{figure} 
\resizebox{\hsize}{!}{\includegraphics{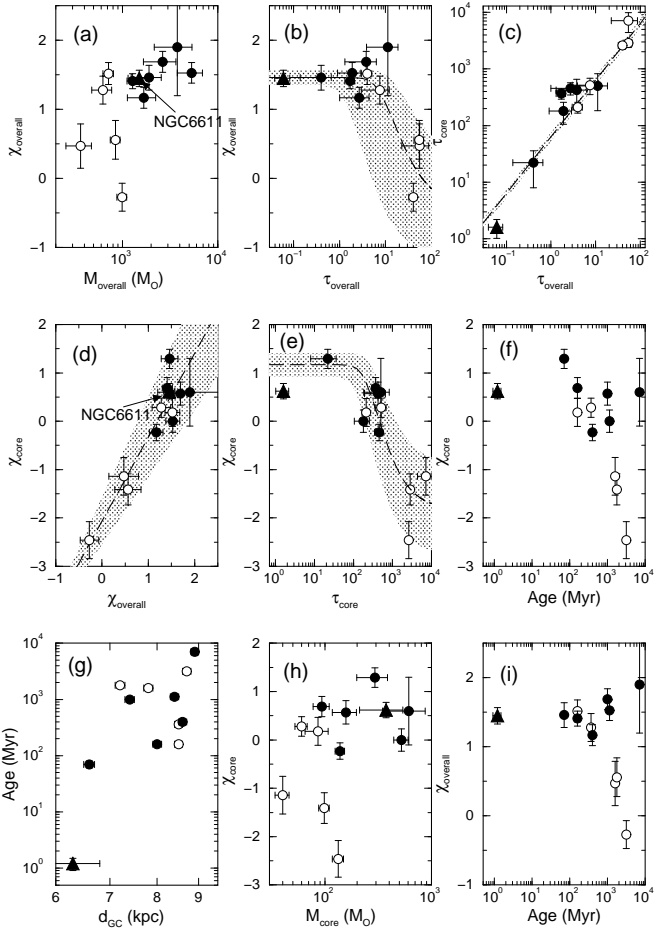}}
\caption[]{Relations involving evolutionary parameters of open clusters. Symbols as in 
Fig.~\ref{fig8}}
\label{fig9}
\end{figure}

NGC\,6611 is compared to older clusters in terms of dynamical-evolution related parameters in 
Fig.~\ref{fig9}. NGC\,6611 fits in the tight correlations of core and overall MF slopes (panel (d); $\rm\chi_{core}=(-2.07\pm0.46)+(1.74\pm0.35)\times\chi_{overall}$, with $\rm CC=0.87$), and core 
and overall evolutionary parameters (panel (c); $\rm\tau_{core}=(63.3\pm9.6)\times\tau_{overall}$, 
with $\rm CC=0.91$). The young age and small Galactocentric distance of NGC\,6611 suggests 
a correlation of \dgc\ with age (panel (g)), which would agree with the result of Lyng\aa\ 
(\cite{Lynga82}). However, we note that this trend is mostly based on a single point (NGC\,6611) 
and thus, a larger cluster sample within the central sector is necessary to establish the statistical 
signficance of the correlation.
The Salpeter-like overall MF slope of NC\,6611 is consistent with the overall mass and age relation
when compared to other massive clusters (panels (a) and (i), respectively). The scatter in 
panel (h) does not allow inferences on the relation of core MF slope with mass. 
Finally, with respect to the evolutionary parameters of NGC\,6611, the overall MF slope (panel (b); 
$\rm\chi_{overall}=(1.5\pm0.1)-(1.9\pm0.6)
\exp{-\left(\frac{16.3\pm13.1}{\tau_{overall}}\right)}$, with $\rm CC=0.89$) 
follows the relations suggested by massive clusters with small values of $\tau$.
However, the core MF slope is flatter than that expected for its value of $\tau$ (panel (e); 
$\rm\chi_{core}=(1.17\pm0.23)-(3.0\pm0.7)
\exp{-\left(\frac{439\pm156}{\tau_{core}}\right)}$, with $\rm CC=0.82$) 
and age (panel (f)).

We conclude that both structurally and dynamically, the overall parameters of NGC\,6611  
fit in the relations defined by older, more-massive open clusters, 
in the loci expected of a very young cluster. 

The core of NGC\,6611 presents a MF significantly flatter than that in 
the halo, a situation typical of older clusters with important mass segregation. Part of this effect
can be accounted for by the fact that the core relaxation time is about half the cluster age.

\section{Concluding remarks}
\label{Conclu}

In this paper we analyzed the structure and mass distribution of MS and PMS stars in the ionizing 
open cluster NGC\,6611. The analysis was based mostly on \jj, \hh\ and \ks\ 2MASS photometry corrected 
for the variable reddening across the region, from which we built radial density profiles and mass 
functions. The average colour excess in a region reaching $\rm R\approx23\arcmin$ from the center of 
NGC\,6611 resulted in $\ejh=0.25$, corresponding to $\ebv=0.8$. To uncover the intrinsic CMD morphology 
we applied a field-star decontamination procedure to the reddening-corrected data. The resulting CMDs 
present a MS restricted to stars more massive than $\approx5\,\ms$ and a collection of PMS stars.

Based on the fraction of stars with \ks\ excess we estimated an age of $\sim1.3\pm0.3$\,Myr for
NGC\,6611. We derived a distance from the Sun of $\rm\ds=1.8\pm0.5\,kpc$, considering the values 
calculated from O\,V stars with previous spectroscopic data and PMS tracks. Our value of \ds\ agrees 
with those given by Tadross (\cite{Tad2002}) and WEBDA. However, it locates NGC\,6611 $\sim0.5$\,kpc 
closer to the Sun than the average value given in previous works. The direct method of distance 
determination using stars is fundamental for the definition of the Galactic structure. Note that 
available kinematic distances of the related \ion{H}{ii} region Sh2-49 are $\ds=2.95$\,kpc (Georgelin, 
\& Georgelin \cite{GG70}) and $\ds=2.2$\,kpc (Blitz, Fich, \& Stark \cite{BFS82}), thus larger than 
the present value. 

King model fits to the radial density profiles including MS and PMS stars produced a core radius 
$\rc=0.70\pm0.08$\,pc with a limiting radius $\rl\approx6.5\pm0.5$\,pc. The projected density of 
stars in the cluster center is $\rm\sim128\pm20\,stars\,pc^{-2}$.

Considering the field star-subtracted LFs, we estimated that the number of PMS stars in NGC\,6611 is 
$\sim362$. The total observed mass locked up in PMS and MS stars amounts to 
$\rm\mtot=(1.6\pm0.3)\times10^3\,\ms$. This is a lower-limit estimate since we are not
taking into account stars less massive than 5\,\ms.

The core MF of the MS stars in NGC\,6611 is flat with a slope $\rm\chi=0.62\pm0.16$, while in the
halo it steepens to $\rm\chi=1.52\pm0.13$. The overall MF of NGC\,6611 is similar to a Salpeter
IMF with $\rm\chi=1.45\pm0.12$. 

Compared to older open clusters in terms of structural and dynamical-evolution-related parameters, 
the overall cluster behaves as a massive open cluster at young ages. 

The spatial variation of MF slopes, being flat in the core and steeper in the halo, implies that 
mass segregation has already affected the mass distribution in the core, since the relaxation time 
$\rm\tr(core)\sim0.73$\,Myr corresponds to $\sim50\%$ of the cluster age. For the whole cluster $\rm\tr(overall)\sim20\,Myr\sim15\times\,$cluster age, which means that mass segregation did not 
have time to redistribute stars in large scale throughout the body of such a young cluster. 
Accordingly, the flattening degree in the core MF of NGC\,6611 seems to be related as well to 
initial conditions, probably associated to the fragmentation of the parent molecular cloud itself, 
with more massive proto-stars preferentially located in the central  parts of the cloud. This agrees 
with Kroupa (\cite{Kroupa2004}) with respect to the observed degree of mass segregation in 
clusters younger than a few Myr.

\begin{acknowledgements}
We are grateful to an anonymous referee for important remarks.
This publication makes use of data products from the Two Micron All Sky Survey, which 
is a joint project of the University of Massachusetts and the Infrared Processing and 
Analysis Center/California Institute of Technology, funded by the National Aeronautics 
and Space Administration and the National Science Foundation. We also made use of the 
WEBDA open cluster database. We acknowledge support from the Brazilian Institutions CNPq
and FAPEMIG.
\end{acknowledgements}

%sssssssssssssssssssssssssssss REFERENCESsssssssssssssssssssssssssssssss
%

\end{document}